\documentclass[intlimits,twoside,a4paper]{article}

\usepackage{amsmath,amssymb}
\usepackage{graphicx}

\usepackage[T2A]{fontenc}
\usepackage[cp1251]{inputenc}

\usepackage[eqsecnum]{cmpj2}
\issue{2016}{19}{4}{43002}
\doinumber{10.5488/CMP.19.43002}

\title[%
Hard spheres in the heat-conduction steady state]%
{Pressure and entropy of hard spheres in the weakly nonequilibrium
heat-conduction steady state}
\author{Y.A. Humenyuk}
\address{Institute for Condensed Matter Physics
of the National Academy of Sciences of Ukraine,\\
1 Svientsitskii St., 79011 Lviv, Ukraine
}

\newcommand{\be}{\begin{equation}}
\newcommand{\ee}{\end{equation}}
\newcommand{\bea}{\begin{eqnarray}}
\newcommand{\eea}{\end{eqnarray}}

\newcommand{\dl}{\mbox{\rm d}}

\newcommand{\eas}[1]{\!\!\! #1 \!\!\!}

\newcommand{\kB}{ k_{\rm B} }

\newcommand{\nn}{\nonumber}
\newcommand{\p}{\partial}

\newcommand{\tsf}[2]{{\textstyle\frac{#1}{#2}}}

\authorcopyright{Y.A. Humenyuk, 2016}
\date{Received April 15, 2016, in final form July 15, 2016}

\begin{document}

\maketitle

\begin{abstract}
Thermodynamic quantities of the hard-sphere system in the steady state with
a small heat flux are calculated within the continuous media approach.
Analytical expressions for pressure, internal energy, and entropy
are found in the approximation of the
fourth order in temperature gradients.
It is shown that the gradient contributions to the internal energy depend on
the volume, while the entropy satisfies the second law of thermodynamics for
nonequilibrium processes.
The calculations are performed for dimensions
3D, 2D, and 1D.

\keywords
heat flux, temperature gradients, equation of state,
nonequilibrium entropy, steady state thermodynamics
\pacs 05.70.Ln, 51.30.+i, 44.10.+i

\end{abstract}

\section{Introduction}

It is known from statistical mechanics that the interparticle interaction
manifests itself in thermodynamic quantities gradually passing from low
gas densities to intermediate ones, e.g.
\cite{75Balescu}.
A similar picture concerning the effect of the interaction on
thermodynamic behaviour should be expected for nonequilibrium states.
They are more complicated and thus are usually investigated for the case
of weak deviations from equilibrium.

As concerns the weakly nonequilibrium states with a heat flux, the main
attention was paid to the phenomenon of heat conduction
\cite{62GroMaz,70Gyarmati,74Zubarev,75Balescu}
as well as to calculations of the linear thermal conductivity coefficient
\cite{70ChaCow,71Silin.E,72FerKap}.
Such nonequilibrium macroscopic quantities as pressure, internal energy,
and entropy have remained less studied.
Interests in the entropy were mainly associated with calculations of its
production
\cite{62GroMaz,70Gyarmati,70ChaCow,71Silin.E,72FerKap,74Zubarev,75Balescu}
closely related to the approaching to equilibrium due to relaxation
processes.

Theoretical investigations of the thermodynamic properties of
systems in the {\em heat-conduction steady\/} state can be divided
into two groups in which a) the effect of the heat flux on the
pressure, entropy, and other quantities and corresponding
densities is studied and b) attempts are made to suggest some
general formalism analogous to the equilibrium Gibbs relation (the
basic thermodynamic equality).
For the hard-sphere model as one of the simplest interparticle
interactions, the Enskog
kinetic equation
\cite{70ChaCow,72FerKap,73BeijerErnst-pu}
is often used in the both cases.

\paragraph{Applications of kinetic theory.}

In order to determine the effect of heat transfer on the weakly
nonequilibrium pressure or entropy, it is necessary to take into account the
terms of higher orders in temperature gradient than the linear ones.
Marques and Kremer
\cite{91MarqueKremer-RevBraFis}
solved the Enskog kinetic equation in the higher approximations using both
Grad's and Chapman-Enskog's methods.
The calculated pressure tensor contains contributions from the temperature
gradients of the second order, which is referred to the Burnett level
\cite{35Burnett-plms}.

The Grad method was also used in solving the Enskog equation for
two-dimensional hard disks
\cite{07UgawaCorder-jsp}.
The set of transport equations of the derived extended hydrodynamics was
applied to two problems:
1)~the first one
\cite{05Ugawa-pa}
was on the pressure difference between the equilibrium and nonequilibrium
stationary heat-conduction hard-disk gases separated by a porous wall;
the phenomenological conclusion on the pressure difference claimed in
\cite{06SasaTasaki-jsp}
had not been confirmed;
\ 2)~the second problem concerned the description of hard disks
between two parallel walls with different temperatures
\cite{07UgawaCorder-jsp};
for the weakly nonequilibrium case, the pressure correction was estimated
to be quadratic in the heat flux.

These results follow from the nonstationary Enskog equation.
In application to the steady case, the pressure calculations were
determined by both stationary and time-dependent parts of the
nonequilibrium distribution function.
Besides, this function was sought as a normal solution, while
the appropriate boundary conditions were not taken into consideration
explicitly. As a consequence, one can deal with local quantities, e.g., the energy and
entropy densities and the intensity of entropy production.
The total energy, entropy, and its production cannot be found,
unless explicit solutions for the hydrodynamic fields of the local number
density and temperature are obtained.

\paragraph{Extensions of the thermodynamic formalism.}

The idea of extension, introduced by Grad
\cite{49Grad-cpam}
for the hydrodynamic level, is known to be applied to the construction of
the
{\em local thermodynamic description\/}
of nonequilibrium states, which is expected to go beyond the domain of the
assumption of local equilibrium.
As a starting equation, the generalized Gibbs-like relation is chosen in the
form written for the local entropy density which depends on the extended
set of hydrodynamic variables
(e.g., including the stress tensor and the heat flux.)

An approach referred to as
{\em Extended Thermodynamics\/}
(developed by
Liu, M{\" u}ller, and Ruggeri with co-workers in works
\cite{72Liu-arma,83LiuMuller-arma,85Liu-arma,89Ruggeri-cmt,%
92SieniuSalamo,98MullerRugger}),
is aimed at analysing the transport equations for an extended set of variables,
to search for the ways of closure procedures for them, and to produce the
criteria of choices of closure relations.
The analysis of the constitutive closure relations is realized on the
grounds of phenomenological principles of invariance of the local
hydrodynamic description, increasing the local entropy, and concavity
of the entropy density functional
\cite{72Liu-arma,83LiuMuller-arma,89Ruggeri-cmt,%
92SieniuSalamo,98MullerRugger}.
However, regardless of the use of entropic
(that is thermodynamic)
criteria, the notion
``extended hydrodynamics'' seems to be more suitable for this approach.

Banach proposed
\cite{87Banach-pa}
microscopic justification of this scheme for the hard sphere
system based on the Enskog kinetic equation of the RET variant
\cite{73BeijerErnst-pu}.
A recent study
\cite{12ArimaTaniguRugger-cmt}
is concerned with the phenomenological analysis and closure procedure for
the extended set of hydrodynamic equations for dense gases, from which the
hard sphere result is derived as a particular case.

{\em Extended Irreversible Thermodynamics\/}
\cite{88JouCasLeb-rpp,99JouCasLeb-rpp,01JouCasLeb-EIT}
develops a formalism of local thermodynamics for nonequilibrium
processes, which does
not exploit the local equilibrium assumption and interprets the
heat flux as an additional thermodynamic degree of freedom.
One expects that it is capable of catching the effects that are
unattainable in the linear irreversible thermodynamics
\cite{62GroMaz,70Gyarmati}.

This approach has been applied to hard spheres
\cite{99RangelVelasc-jms}.
Using the Enskog kinetic equation and the method of molecular hydrodynamics
\cite{80BoonYip},
explicit expressions for the phenomenological coefficients appearing in the
extended Gibbs relation are calculated.

However, the extended irreversible thermodynamics and specifically its
notion of the nonequilibrium temperature
\cite{94CasasJou-pre,03CasasJou-rpp}
were criticized from the viewpoint of computer simulations
\cite{93HooverHolianPosch-pre,08HooverHoover-pre},
using phenomenological ideas
\cite{95Garcia-mp,98BhalekGarcia-prmn},
and even on the grounds of its own internal methodology
\cite{93Henjes-pre}.
To our knowledge, the discussion has not been resolved as well as no
crucial experiment which would approve or refute the main assumptions of
this approach has been proposed.

\paragraph{Computer simulations.}

The hard spheres or disks are investigated in computer simulations with
regard to their properties in the heat-conduction state and behaviour of
local quantities.
The measure of deviation of the results for the heat flux from the linear Fourier
law is also studied.

Using the nonequilibrium molecular dynamics, the temperature and number
density profiles for the hard-disk system are obtained
\cite{96RissoCorder-jsp},
which are shown to be in agreement with the results of the continuum media
approach.
However, the thermal conductivity coefficient is found
\cite{96RissoCorder-jsp}
to differ appreciably from that given by the Enskog theory.

The steady states of hard spheres with a heat flux are explored by
numerical methods of solving the Enskog equation.
One of them is the generalization
\cite{99Frezzotti-ejmb}
of the direct simulation Monte Carlo,
proposed and developed by Bird
\cite{63Bird-pf,70Bird-pf,94Bird-DSMC}.
In
\cite{15WuZhangReese-JCompPhy},
the spectral method is used for solving the Enskog equation for hard spheres
(both elastic and inelastic) in the heat-conduction states.
The profiles of the number density, kinetic as well as potential components of
the pressure and heat flux are shown to agree well with the direct
simulation Monte Carlo data
\cite{99Frezzotti-ejmb}.

Morriss with co-workers consider a simplified spatial configuration
---
the quasi-one-dimensional system of hard disks in a narrow linear channel with
model thermal baths on the ends
\cite{03TaniguMorris-pre}.
The disks are coupled to the thermostats by deterministic rules
\cite{03TaniguMorris-pre,08MorrisChungAngstm-Ent,12MorrisTruant-Ent}.
The temperature profile, the local entropy density, its production, and
the heat flux through the system are obtained for both low
\cite{09KimMorris-pre}
and intermediate and high
\cite{13MorrisTruant-pre}
densities.
The effect of spatial correlations on the local entropy is examined in
\cite{14Morriss-cm}.

These numerical methods provide results describing the
heat-conduction steady states in detail.
However, they do not solve the problem of establishing theoretical
interrelations between different macroscopic quantities.
In recent works by del Pozo~et~al.
\cite{15PozoGarridHurtad-pre.Scaling,15PozoGarridHurtad-pre.Probing}
computer simulation data for the two-dimensional hard disks in the
heat-conduction steady states are analyzed in terms of the
equilibrium-like equation of state and the local Fourier law.
Bulk behaviour of the temperature and particle density profiles
are shown to obey specific scaling relations valid even for strong
nonequilibrium conditions.
High accuracy and reliability of these objective laws considerably deepen
the understanding of the nature of the steady states.

In
\cite{15H-ujp},
there is calculated the pressure, internal energy, entropy, and free energy
(not accurately)
of the low-density gas in the weakly nonequilibrium heat-conduction steady
state by means of the continuous media approach.
Simplicity of the method and the fact that the entropy found satisfies the
second law of thermodynamics show the usefulness of these results.
However, an interaction potential does not enter the thermodynamic
quantities with regard to low densities.
Here, we attempt to take interparticle interaction into account for the
particular case of the hard-sphere system at intermediate densities making
use of one of its simplest equations of state.
This demonstrates the applicability of the method to the calculation of
thermodynamic quantities of gases in the situations where the size of
particles becomes important.

In
section~\ref{2tPSStK} we describe the heat-conduction steady state.
Next, we find the pressure and internal energy,
section~\ref{3rsPE}.
The entropy calculations and conclusions are given in
section~\ref{4EntViE}
and
section~\ref{Pidsumky}.

\section{Heat-conduction state of the hard spheres}
\label{2tPSStK}

\begin{figure}[!t]
\begin{minipage}{0.49\linewidth}
\begin{center}
\includegraphics*[bb=59 491 559 800,width=1.0\hsize]%
{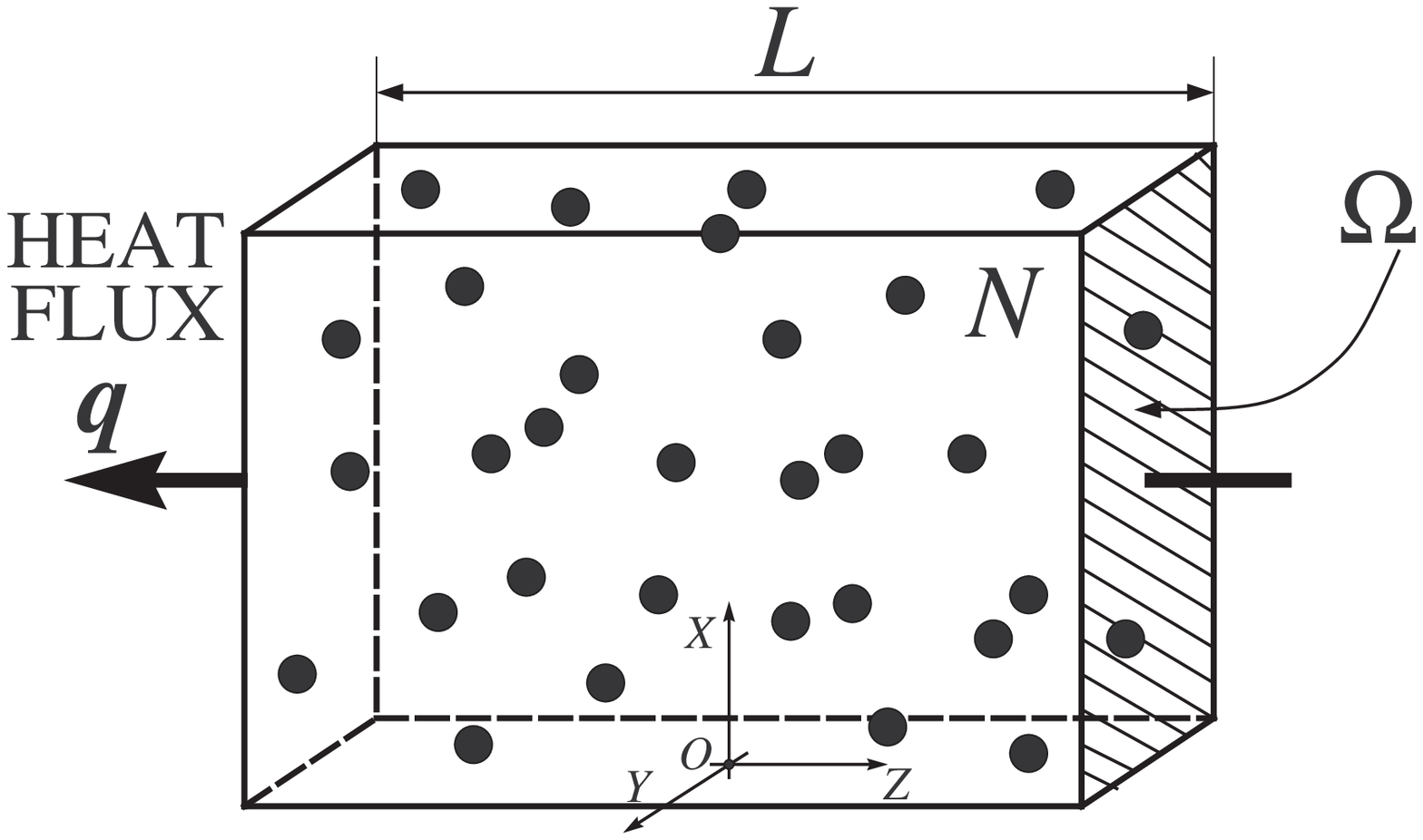}
\end{center}
\caption{\label{piped} The hard-sphere system in the heat-conduction steady state.}
\end{minipage}
\begin{minipage}{0.49\linewidth}
\begin{center}
\includegraphics*[bb=53 398 583 775,width=0.9\hsize]%
{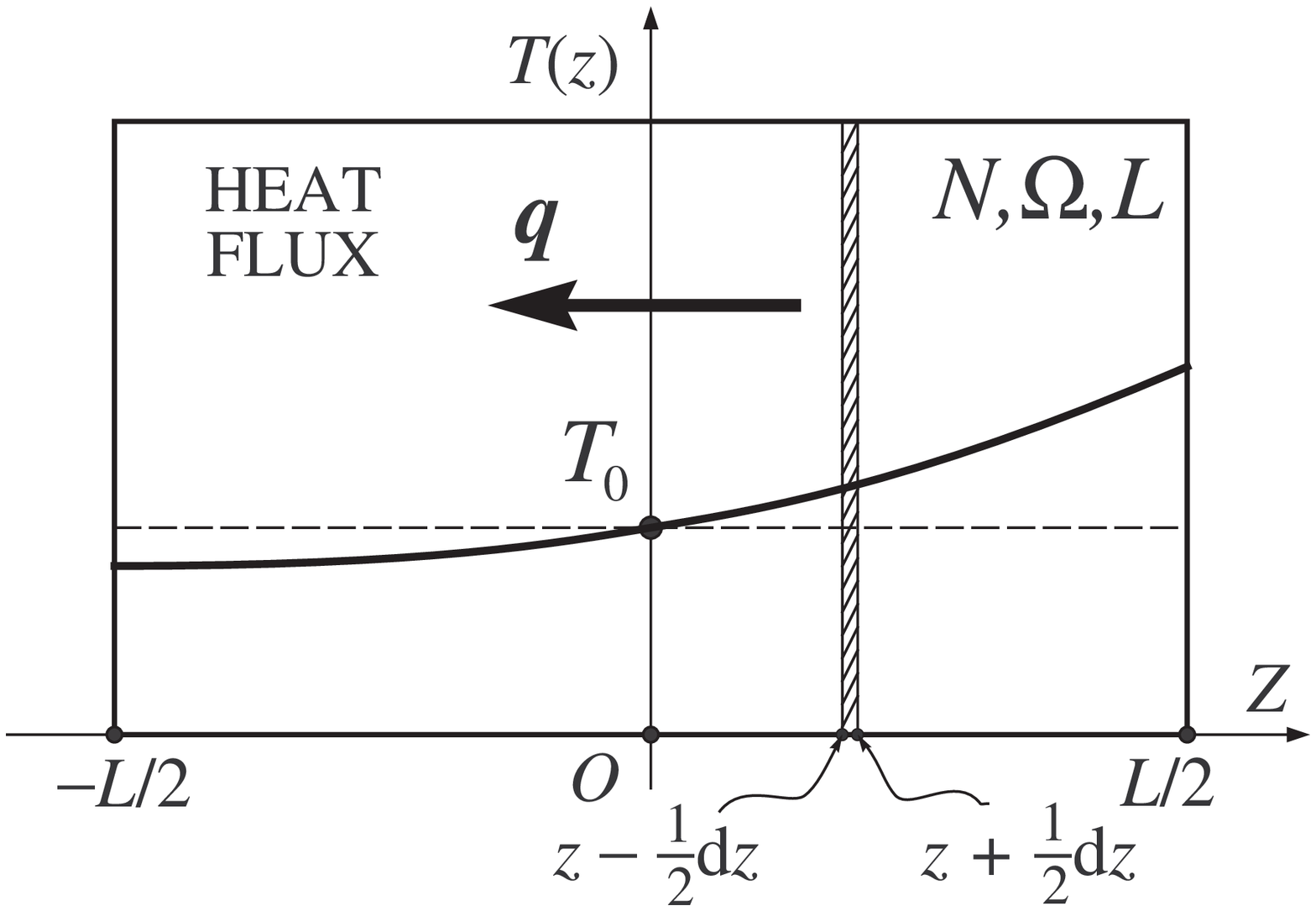}
\end{center}
\vspace{-2mm}
\caption{\label{xidT} The local temperature profile along the hard-sphere system.}
\end{minipage}
\end{figure}

Our aim is to study the effect of the size of molecules on
thermodynamic quantities of the intermediate density gas in the
{\em heat-conduction steady state\/}
using the hard-sphere model.
We restrict ourselves to a simple case of
{\em weak nonequilibrium.}
$N$ hard spheres are contained in a vessel of macroscopic size and of a
parallelepiped form.
The length of the edge and the cross area are denoted by
$L$
and
$\Omega$
(figure~\ref{piped}).
Heat is transferred in the direction parallel to the edge, while the
local temperature is independent of time and changes
{\em slowly\/} along this direction.

\paragraph{Local temperature.}

Putting the explicit determination of the temperature profile off,
we consider the problem of calculation of the thermodynamic
quantities from rather general grounds and as before
\cite{15H-ujp} we
describe the steady state by the set of temperature value
$T_0$ and values
$\{ G_1, \ldots, G_r\}$
of its
$r$ successive gradients referred to the
{\em geometrical middle-point\/} of the vessel.
If axis
$OZ$
of the reference system is chosen to be parallel to
the heat flux
(figure~\ref{xidT}),
then the quantities
$$
T_0 \equiv T(z) \Big|_{z=0} \,,
\qquad
G_k \equiv \frac { \p^k }{ \p z^k } T(z) \Big|_{z=0}
$$
can approximately determine the local temperature:
\be
\label{T.z:G}
T(z) = T_0  +
G_1 z + \frac 1{2!} G_2 z^2 + \ldots + \frac 1{r!} G_r z^r
.\ee
The approximation is defined by the number of the gradients in
equation~(\ref{T.z:G}).

The weak nonequilibrium means that any two neighbouring terms
in equation~(\ref{T.z:G}) differ by an order
$ \tsf 1 k| G_{k+1} z | \ll | G_k |
$
for all
$z$.
For boundary values
$z = \pm L/2$,
these inequalities read:
\be
\label{u-G.k}
\frac 1 {2(k+1)}
| G_{k+1} |  L  \ll |G_k|
,\ee
in particular,
$
\tsf 12 |G_1| L  \!\ll\! T_0
$,
$\,
\tsf 14
|G_2| L  \!\ll\! |G_1|
$,
etc.
Such conditions are standard in nonequilibrium statistical
mechanics and kinetic theory,
e.g.,
\cite{62GroMaz,70ChaCow,74Zubarev}.
It is convenient to distinguish different orders
by a formal small parameter
$\delta$
introduced into expansion
(\ref{T.z:G}):
\be
\label{T.z:g}
T(z) = T_0  \big[ 1 + \delta g_1 z + \ldots + \delta^r g_r z^r \big]
,\ee
where reduced gradients
$\{ g \} \equiv \{ g_1, \ldots, g_r\}
$
defined by
$g_k  \equiv \frac 1 {T_0}  \frac 1 {k!} G_k
$
are used instead of 
$\{ G \} \equiv \{ G_1, \ldots, G_r\}
$.
According to this
$\delta$-expansion,
any macroscopic quantity
$A$ will be represented below as a series in powers of
$\delta$:
\be
\label{A-R}
A  = A_0  + \sum_{i=1}^r \delta^i  A_i,
\ee
where
$A_i$ contains contributions from the gradient combinations of order
$i$.

\paragraph{Local equation of state.}

If we select
(figure~\ref{xidT}) the
{\em macroscopically small\/} layer
$[z - \tsf 12 \dl z ; z + \tsf 12 \dl z ]$
(but sufficiently large in comparison with the hard-sphere diameter),
then with regard to the weak nonequilibrium of the state, the pressure in this
layer can be
{\em approximated\/}
by the equilibrium equation of state.
We choose the latter to be the van der Waals equation for hard spheres,
e.g.,~\cite{75Balescu,72Kobyljans.E}:
\be
\label{P.VdW}
P_{V\rd W-HS} =
\frac {N \kB T} {V - Nb}\,
,\ee
where
$V$ is the volume of an equilibrium system and
$b$
means the volume referred to a particle in the close-packing state.
The corresponding internal energy and entropy read:
\bea
\label{E.tK-o}
E_{\rm eq} &\eas{\equiv}&  \frac 12 D N \kB T ,
\\
\label{S.tK-o}
S_{\rm eq} &\eas{\equiv}&
N \kB
\left[ - \ln N + \ln(V- Nb)  +\frac 12 D \ln T  +\xi_S^{(D)}
\right] ,
\eea
with
$D$ being the dimensionality and
$\xi_S^{(D)} \equiv \tsf 12 D \ln ( 2\pi \kB m/h^2) +1 +\tsf12 D$,
where
$m$ is the mass of the particle and
$h$ is Planck's constant,
see e.g.,~\cite{75Balescu}.

We substitute the
{\em real local values\/} of the temperature
$T(z)$ and the number density
$n(z)$ into
equation~(\ref{P.VdW})
to get the
{\em local pressure assumption\/}
for the weakly nonequilibrium heat-conduction steady state:
\be
\label{P.z-VdW}
P(z) =
\frac {n(z) \kB T(z)} {1 - b n(z)}\,.
\ee
The local densities of the internal energy and entropy can be obtained from
the equilibrium counterparts
$E_{\rm eq}/V$
and
$S_{\rm eq}/V$
in the same way:
\bea
\label{veps.z-o}
\varepsilon (z) &\eas{\equiv}&  \frac 12 D \, n(z) \kB T(z)
,\\
\label{s.z-o}
s (z) &\eas{\equiv}&
\kB n(z)
\left(
\ln \big\{ [n(z)]^{-1} - b \big\}  +\frac 12 D \ln T(z)  +\xi_S^{(D)}
\right)
.\eea

\section{Baric and caloric equations of state}
\label{3rsPE}

Next, we turn to calculation of the pressure.
The fact that the hard spheres are maintained in a
{\em mechanical\/}
equilibrium means that the pressure has the same value all
over the vessel:
\be
\label{P=const}
P (z) = {\rm const}.
\ee
This statement is a natural condition for the heat-conduction steady state.
It is involved in the statistical-mechanical description of light scattering
\cite{79KirkpaCohenDorfma-prl.KiT,80KirkpaCohenDorfma-prl.HD,86Morozov-tmf.E}
and the BGK-model kinetic calculations
\cite{87BreySantosDufty-pra.HeaMomTra,89SantosBreyKimDufty-pra.VelDis},
while in computer simulations checking of this condition ensures
additionally the validity of results for the steady state,
e.g.,
\cite{09KimMorris-pre}.
Consequently, we have
{\em a constant quantity\/} in the left-hand side of
equation~(\ref{P.z-VdW})
for the case of the steady state,
henceforth denoted as
$P$.

\paragraph{Number density}

$n(z)$ obeys the normalization condition:
\be
\label{uNor-n.z}
\Omega \int_{-L/2}^{L/2} \!\dl z \: n(z) = N ,
\ee
where integrations with respect to transverse coordinates
$x$ and $y$
have been performed%
\footnote{
\label{Z-Omega}
$\Omega = 1$
in the
1D case, while for the
2D case
$\Omega$ means the
{\em linear size\/} in the perpendicular direction.
}
in the integral over the volume
$\Omega \!\times\! L$;
$N$ is the total number of particles in the system.
The density
$n(z)$ can be expressed through
$T(z)$ and
$P$ using
equation~(\ref{P.z-VdW}):
$
n(z) =  C /[T(z) + bC ]
,$
where
$C \equiv P/\kB$ is a constant.
Substitution of the expansion for the local temperature,
equation~(\ref{T.z:g}), yields:
\be
\label{n.z:gamma}
n(z) = n_0 \,\frac 1 { 1 + \delta \gamma_1 z + \ldots + \delta^r \gamma_r z^r } \,,
\ee
with
\be
\label{n.0-o}
n_0 \equiv \frac C { T + bC } \,,
\qquad
\gamma_k \equiv \frac 1 {k!} \,\frac 1 { T + bC } \,G_k ;
\ee
here and below, the middle-point temperature value is denoted by
$T$
(in place of $T_0$).

It follows from the weak nonequilibrium conditions
(\ref{u-G.k})
that
$\sum \delta^k \gamma_k z^k \ll 1$
and the fraction in
equation~(\ref{n.z:gamma}) can be expanded
(up to the
$r$-th order):
\be
\label{n.z-R}
n(z) = n_0 \big[ \nu_0 + \delta \nu_1 z + \ldots + \delta^r \nu_r z^r + \ldots \big]
,\ee
with coefficients dependent on
$C$
through the parameters
$\{ \gamma \}$,
equation~(\ref{n.0-o}):
$$
\label{nu.1-4}
\begin{array}{ll}
\nu_0 = 1
,&
\nu_3 (C) =  -\gamma_3 + 2 \gamma_2 \gamma_1 - \gamma_1^3
,\vspace{2mm}\\
\nu_1 (C) = -\gamma_1
,&
\nu_4 (C) = -\gamma_4 + 2 \gamma_3 \gamma_1 + \gamma_2^2
- 3 \gamma_2 \gamma_1^2 + \gamma_1^4
,\vspace{2mm}\\
\nu_2 (C) = -\gamma_2 + \gamma_1^2
,\qquad\qquad&
\ldots
\end{array}
$$
In what follows, we restrict ourselves to the fourth order though
it is not so hard to derive higher-order contributions.

\paragraph{Perturbations for the pressure.}

Equation
(\ref{n.z-R}) inserted into the normalization condition
(\ref{uNor-n.z}) can be integrated explicitly
\be
\label{eq-C}
\overline{n} ( T + bC )  = C
\big[ 1  + \delta^2 \varkappa_2 (C)
+ \delta^4 \varkappa_4 (C) + \ldots \big]
,\ee
where
equation~(\ref{n.0-o}) has been used for
$n_0$;
here,
$\overline{n} \equiv N/ \Omega L$
is the number density in the state of thermal equilibrium, while the
coefficients introduced read:
\be
\label{vkap-nu}
\varkappa_2 (C) \equiv \frac 1{12} \nu_2(C) L^2
,\qquad
\varkappa_4 (C) \equiv \frac 1{80} \nu_4(C) L^4
.\ee
Expression
(\ref{eq-C})
is an equation to determine the constant
$C$.
Since
$\nu_2$, $\nu_4,\,\ldots$
depend on
$C$ too,
equation~(\ref{eq-C}) is highly nonlinear.
Its solution for the weak nonequilibrium can be sought by
perturbations:
\be
\label{C-R}
C = C^{(0)} + \delta^2 C^{(2)} + \delta^4 C^{(4)} + \ldots\,.
\ee
Even orders in
$\delta$ are absent here because they are absent in
equation~(\ref{eq-C}).

We note that the coefficients
$\varkappa_2$, $\varkappa_4, \,\ldots$
are also to be expanded in
$\delta$:
\be
\label{vkap.i-R}
\varkappa_i (C) = \varkappa_i^{(0)}
+ \delta^2 \varkappa_i^{(2)} + \delta^4 \varkappa_i^{(4)} + \ldots,
\qquad i = 2; \,4; \,\ldots ,
\ee
where
$\varkappa_i^{(k)}$ is caused by those contributions to
$C$
whose order is
{\em not higher\/} than
$k$;
in particular,
$\varkappa_i^{(0)}$
is determined by the term
$C^{(0)}$,
$\varkappa_i^{(2)}$
is determined by the terms
$C^{(0)}$ and $C^{(2)}$, etc.
After substitution of expansion
(\ref{vkap.i-R}) into
equation~(\ref{eq-C}), the series in the square brackets is rearranged:
\be
\label{pHrup-vkap}
1  +  \delta^2 \varkappa_2  + \delta^4 \varkappa_4 + \ldots
=
1  + \delta^2 \varkappa_{(2)} + \delta^4 \varkappa_{(4)} + \ldots
,\ee
where
$\varkappa_{(2)} \equiv \varkappa_2^{(0)}$
and
$\varkappa_{(4)} \equiv \varkappa_4^{(0)} + \varkappa_2^{(2)}$
contain terms of their own orders in the gradients.
In accordance with
equations~(\ref{vkap-nu}),
they are related to
\be
\label{nu-pH-vkap}
\nu_{(2)} \equiv \nu_2^{(0)} ,
\qquad
\nu_{(4)} \equiv \nu_4^{(0)} + \frac {20}3 L^{-2} \nu_2^{(2)} ,
\qquad
\ldots
\ee
Explicit expressions for these coefficients are given in the Appendix.

Insertion of
equations~(\ref{C-R}) and (\ref{pHrup-vkap}) into expression
(\ref{eq-C})
gives equations for
$C^{(k)}$:
$$
\overline{n} \,T + \eta C^{(0)} = C^{(0)} ,
\qquad\;
\eta C^{(2)} = C^{(0)} \varkappa_{(2)} + C^{(2)} ,
\qquad\;
\eta C^{(4)} = C^{(0)} \varkappa_{(4)} + C^{(2)} \varkappa_{(2)} + C^{(4)}
,$$
with
$
\eta \equiv \overline{n} b
$
being the reduced partial volume.
Finally, we obtain the solutions:
\bea
&& \nn
C^{(0)} = \overline{n} \,T \,\tilde\eta^{-1} ,
\\
&& \nn
C^{(2)} = C^{(0)} \tilde\eta^{-1} [ - \varkappa_{(2)} ] ,
\\
&& \nn
C^{(4)} = C^{(0)} \tilde\eta^{-1}
\big[ - \varkappa_{(4)}  + \tilde\eta^{-1} \varkappa_{(2)}^2
\big]
.\eea
Here,
$
\tilde\eta \equiv 1 - \eta
$
denotes the reduced accessible volume.
The result for
$C^{(4)}$
has been obtained by the use of the formula for
$C^{(2)}$.

The contributions
$C^{(k)}$ can be expressed through the gradients
$\{ g \}$
and we deduce the baric equation of the weakly nonequilibrium
heat-conduction steady state for hard spheres in the van der Waals
approximation
\cite{16H-icmp.tK.E}:
\be
\label{P-r}
P (N,\Omega,L; T, g_1, \ldots, g_4)
= \frac { N \kB T } { \Omega L - N b}
\,( p_0 + p_2 + p_4 + \ldots
)
,\ee
where powers of
$\delta$ are omitted, and
$p_0 \equiv 1
$,
$p_2 \equiv  \tsf 1 {12}
\big( g_2  - g_1^2 \tilde\eta \big) L^2
$, and
$$
p_4 \equiv  \frac 1 {80}
\left[
g_4 - 2 g_3 g_1 \tilde\eta
- \frac49 g_2^2 \tilde\eta
+ \frac {17} 9 g_2 g_1^2 \tilde\eta \left( 1 - \frac {12}{17} \eta \right)
-\frac49 g_1^4 \tilde\eta^2 \left( 1 + \frac 14 \eta \right)
\right] L^4
.$$
The quantities
$p_2$ and $p_4$
describe the corrections to the pressure from the gradients in corresponding
orders.
The effect of particle's size is involved in the reduced volume
$\eta$
referred to the particles, mainly in combination
$\tilde\eta \equiv 1-\eta$.
Tending
$b \to 0$ causes
$\eta \to 0$,
$\tilde\eta \to 1$
which results in the expressions transforming to those for the
low-density case
\cite{15H-ujp}.

The gradient expansion for the middle-point value of the number density,
equation~(\ref{n.0-o}), can be also found as
\be
\label{n.0-R}
n_0 = n_0^{(0)} + n_0^{(2)} + n_0^{(4)} + \ldots
,\ee
in which the coefficients are defined as follows
\cite{16H-icmp.tK.E}:
$n_0^{(0)} = \overline n
$,
$n_0^{(2)} = \overline n \tsf 1 {12}
\big( g_2 \tilde\eta - g_1^2 \tilde\eta^2 \big) L^2
$, and
$$
n_0^{(4)} = \overline n
\frac 1 {80}
\left[
g_4 \tilde\eta - 2 g_3 g_1 \tilde\eta^2
- \frac49 g_2^2 \tilde\eta \left( 1 + \frac 14 \eta \right)
+ \frac {17} 9 g_2 g_1^2 \tilde\eta^2 \left( 1 - \frac 2{17} \eta \right)
-\frac49 g_1^4 \tilde\eta^3 \left( 1 + \frac 32 \eta \right)
\right] L^4
.$$

\paragraph{The internal energy \hspace{-0.5em}}

is calculated by integration of its density
$\varepsilon (z)$,
equation~(\ref{veps.z-o}):
\be
\label{E-o}
E \equiv
\Omega \int_{-L/2}^{L/2}  \! \dl z \: \varepsilon (z) .
\ee
It follows from the local equation of state
(\ref{P.z-VdW}) that
$
\varepsilon (z) =
\tsf 12 D P [ 1 - b n(z)  ]
$
is not constant along the heat flux, contrary to the low-density gas case
\cite{15H-ujp}.
Normalization condition
(\ref{uNor-n.z})
simplifies the integration in
equation~(\ref{E-o}) resulting in
$
E =
\tsf 12 D P \Omega L \tilde\eta
.$
Using expression for
$P$ yields:
\be
\label{E-r}
E (N,\Omega,L; T, g_1, \ldots, g_4)
=  \frac 12 D N \kB T
(  1 + e_2 + e_4 + \ldots  )
\ee
with coefficients
$e_k = p_k$ dependent on
$\{g\}$
and
$\eta = b N/(\Omega L)$.
We conclude that the internal energy of hard spheres in the
heat-conduction steady state
{\em depends on the volume}
$\Omega L$
and differs from the low-density result
\cite{15H-ujp},
while the equilibrium energies of these systems are known to be identical and
independent on volume.

\section{Entropy}
\label{4EntViE}

Expression
(\ref{s.z-o}) for
$s(z)$ is related to a number of states in the phase space owing to the
starting equilibrium entropy
(\ref{S.tK-o}).
For this reason, we accept an integral quantity
\be
\label{SF-o}
S \equiv
\Omega \int_{-L/2}^{L/2}  \! \rd z
\, s(z)
\ee
to be the entropy of the weakly nonequilibrium heat-conduction steady
state.
After its calculation,
$S$ is shown to satisfy the second law of thermodynamics for
nonequilibrium processes
\cite{62GroMaz,70Gyarmati,83Bazarov.E}.

\paragraph{Contributions to $S$.}

Using the local equation of state
(\ref{P.z-VdW})
for the expression
$\big\{ [n(z)]^{-1} - b \big\}$
in
equation~(\ref{s.z-o}),
we obtain:
\be
\label{s.z+f.z}
s(z) =
\kB n(z)
\Big[ d_1 \ln T(z) - \ln ( P/\kB ) +
\xi_S^{(D)} \Big]
,\ee
with
$d_1 \equiv \tsf 12 D +1$.
This expression inserted into
equation~(\ref{SF-o})
produces three contributions corresponding to the terms
in the square brackets:
$$
S = S_T + S_P + S_\xi\,
.$$

$S_P$ and $S_\xi$
can be found by virtue of the normalization condition,
equation~(\ref{uNor-n.z}):
\be
\label{S.P}
S_P = - N \kB \ln (P /\kB)
,\qquad
S_\xi = N \kB \xi_S^{(D)}
.\ee
Expanding
$\ln (P /\kB)$ into a series yields
\be
\label{S.P-R}
S_P = N \kB
\big( s_{P,\,0}^{} + s_{P,\,2}^{} + s_{P,\,4}^{} + \ldots
\big)
,\ee
where the coefficients read:
$
s_{P,\,0}^{}
\equiv - \ln ( \overline n \,T \tilde\eta^{-1} )
$,
$
s_{P,\,2}^{}
\equiv \tsf 1{12} \big( - g_2 + g_1^2 \tilde\eta \big) L^2
$,
and
$$
s_{P,\,4}^{}
\!\equiv\! \frac 1{80} \left[
- g_4
+ g_3 g_1 2 \tilde\eta
+  g_2^2 \left(\frac {13}{18} - \frac 49 \eta\right)
+ g_2 g_1^2 \tilde\eta \left(-\frac {22} 9 + \frac 43 \eta\right)
+ g_1^4 \tilde\eta^2 \left(\frac {13}{18} + \frac 19 \eta\right)
\right] L^4
.$$

\paragraph{Calculation of $S_T$.}

We insert
equation~(\ref{s.z+f.z}) into definition
(\ref{SF-o}):
\be
\label{SF.T-o}
S_T
\equiv
\Omega d_1
\kB
\int_{-L/2}^{L/2}  \! \rd z
\, w(z)
,\ee
where
$w(z) \equiv n(z) \ln T(z)$.
After the expansion of this function is used,
$$
w(z) = n_0 \left( w_0 + w_1 z + w_2 z^2 + \ldots \right)
,$$
we can integrate in
equation~(\ref{SF.T-o}) in explicit way:
\be
\label{S.T:w.i}
S_T
\,=\;
\Omega d_1 \kB n_0 L \left(
w_0 + \frac 1{12} w_2 L^2 + \frac 1{80} w_4 L^4 + \ldots \right) .
\ee

Now, we use the expansion
(\ref{n.z-R})
for
$n(z)$ and the series
$
\ln T(z) = \tau_0  +\tau_1 z  +\tau_2 z^2  +\ldots
$
with the coefficients calculated from
equation~(\ref{T.z:G}), which read:
$$
\begin{array}{ll}
\label{tau.0-4}
\tau_0 = \ln T
,&
\tau_3 = g_3 - g_2 g_1 + \frac13 g_1^3
, \vspace{2mm}\\
\nn
\tau_1 = g_1
,&
\tau_4 = g_4 - g_3 g_1 - \frac12 g_2^2 + g_2 g_1^2 - \frac14 g_1^4
,\vspace{2mm}\\
\nn
\tau_2 = g_2 - \frac12 g_1^2
,\qquad\qquad&
\ldots
\end{array}
$$
The quantities
$w_i$
can be identified as discrete convolutions of
$\{ \nu \}$ and
$\{ \tau \}$:
\be
\label{w.i=nu*tau}
w_i \equiv \nu_i \tau_0 +  \ldots +
\nu_{i-k} \tau_k + \ldots
+ \nu_0 \tau_i
.\ee

The series in
equation~(\ref{S.T:w.i})
is to be rearranged in a similar way to the pressure case given above,
as
$w_i$
are linear combinations of coefficients
$\{ \nu \}$:
\be
\label{S.T:w.(i)}
S_T = \Omega d_1 \kB n_0 L
\left[
w_{(0)} + \frac 1{12} w_{(2)} L^2 + \frac 1{80} w_{(4)} L^4 + \ldots
\right]
,\ee
where the new coefficients read:
$$
w_{(0)} \equiv w_0 ,
\qquad\qquad
w_{(2)} \equiv w_2^{(0)} ,
\qquad\qquad
w_{(4)} \equiv w_4^{(0)} + \frac {20}3 L^{-2} w_2^{(2)} ,
\qquad
\ldots
;$$
here,
$
w_i^{(0)} \equiv \sum\nolimits_{k=0}^i \nu_{i-k}^{(0)} \tau_k
$
and
$
w_2^{(2)} \equiv \nu_2^{(2)} \tau_0 + \nu_1^{(2)} \tau_1
$.

After substitution of expansion
(\ref{n.0-R}) for
$n_0$
into
equation~(\ref{S.T:w.(i)}),
we obtain the following result:
\be
\label{S.T-R}
S_T = N \kB
\big( s_{T,\,0}^{} + s_{T,\,2}^{} + s_{T,\,4}^{} + \ldots
\big)
,\ee
where
$s_{T,\,0}^{} \equiv d_1 \ln T
$,
$s_{T,\,2}^{}
\equiv d_1 \frac 1{12} \big[ g_2 - g_1^2 \left(\tsf 32 - \eta\right) \big] L^2
$,
and
\bea
\nn
s_{T,\,4}^{}
& \eas\equiv& d_1 \frac 1{2^6 \!\cdot\! 3^2 \!\cdot\! 5} \big[
36 g_4
+ (-108 + 72 \eta) g_3 g_1
+ (-34 + 16 \eta) g_2^2
+{}
\quad
\\
\nn
& \eas{+}&
\left(148 - 160 \eta + 48 \eta^2\right) g_2 g_1^2
+ \left(-45 + 56 \eta - 16 \eta^2 - 4 \eta^3\right) g_1^4
\big] L^4
.\eea

\paragraph{Compatibility with the second law of thermodynamics.}

We collect the contributions found in
equations~(\ref{S.P}), (\ref{S.P-R}), and (\ref{S.T-R})
to obtain the final expression
\cite{16H-icmp.tK.E}:
\be
\label{rez-S}
S = N \kB
\left( s_0^{} + s_2^{} + s_4^{} + \ldots
\right)
\ee
with the coefficients defined by
$s_i \equiv s_{T,i} + s_{P,i} + s_{\xi,i}$:
\bea
\nn
s_0^{}
& \eas\equiv&
\ln (\Omega L/N - b)
+ \frac D2 \ln T + \xi_S^{(D)}
,\\
\nn
s_2^{}
& \eas\equiv& \frac 1{2^3 \!\cdot\! 3} \left\{
D g_2 + \left[ \left(-\frac 32 + \eta\right) D - 1 \right] g_1^2 \right\} L^2
,\\
\nn
s_4^{}
& \eas\equiv& \frac 1{2^7 \!\cdot\! 3^2 \!\cdot\! 5} \big(
\sigma_4 g_4
+ \sigma_{31} g_3 g_1
+ \sigma_2 g_2^2
+ \sigma_{21} g_2 g_1^2
+ \sigma_1 g_1^4
\big) L^4
;\eea
the
$\sigma_\alpha$'s
multiplying the gradients in
$s_4^{}$
are written as follows:
$$
\begin{array} {ll}
\!\sigma_4 \equiv 36 D
,&
\sigma_{21} \equiv (148 - \!160 \eta\! + 48 \eta^2) D\! + \!120 - 48 \eta
,\\
\!\sigma_{31} \equiv (-108 + 72 \eta) D - 72
,\qquad\quad&
\sigma_1 \equiv (-45 + 56 \eta - 16 \eta^2 - 4 \eta^3) D- 38 + 16 \eta + 4 \eta^2
.
\\
\!\sigma_2 \equiv (-34 + 16 \eta) D - 16
,&
\qquad {}\end{array}
$$
We notice that the coefficients
$s_i$
for the entropy
(an additive quantity)
depend on the dimensionality
$D$,
contrary to the pressure ones,
$p_i$.
In the limit
$b \to 0$
(and $\eta \to 0$),
the low-density gas results are recovered, which coincide for value
$D=3$
with those found earlier
\cite{15H-ujp}.

Any nonequilibrium state undergoes a relaxation at the conditions of
the free evolution, which is accompanied by the
{\em entropy increase\/}
\cite{62GroMaz,70Gyarmati,83Bazarov.E}.
We show that the entropy calculated
(\ref{rez-S})
possesses this feature.
To this end, let us imagine that the system is made isolated on the boundaries
$z = \mp \tsf 12 L$
and afterwards it is allowed to relax during a macroscopically large
time interval.
The entropy
$S_{\rm fin}$
of the final equilibrium can be compared to that of the initial steady
state,
equation~(\ref{rez-S}).

The internal energy of the hard spheres does not change after the
isolation, thus
$E = \tsf D2 N \kB T_{\rm fin}$,
where
$T_{\rm fin}$
is the temperature ascribed to the final state.
We derive from
equation~(\ref{E-r})
for the internal energy:
$$
T_{\rm fin} =  T \big(  1 + e_2 + e_4 + \ldots  \big) .
$$
Substitution of this result into
equation~(\ref{E.tK-o})
leads to the entropy of the final equilibrium:
$$
S_{\rm fin} =
N \kB
\left\{  \ln \left( \frac {\Omega L} N - b\right)  + \frac D2 \ln T
+\frac D2 \left[ p_2  + \left(p_4 - \frac 12 p_2^2 \right) + \ldots \right]
+\xi_S^{(D)}
\right\} ,
$$
where we have used that
$e_k = p_k$,
while the third term in the curly brackets is an expansion of the
logarithm.

The entropy difference
$\Delta S \equiv S - S_{\rm fin}$
takes the form
$\Delta S = \Delta S^{(2)} + \Delta S^{(4)} + \ldots$
with
\bea
\nn
\Delta S^{(2)} &\eas\equiv& N \kB
\frac {D+2} {48}
\big(
- g_1^2 L^2
\big)
,\\
\nn
\Delta S^{(4)} &\eas\equiv& N \kB
\frac {D+2} {2^7 \cdot 3^2 \cdot 5}
\big[
-36 g_3 g_1 -8 g_2^2 + (60 - 24 \eta) g_2 g_1^2
+ \left(-19 + 8 \eta + 2 \eta^2\right) g_1^4
\big] L^4
.\eea
It is obvious that
$\Delta S^{(2)} <0$,
while the sign of
$\Delta S^{(4)}$
is undetermined and depends on the values of the \linebreak fourth-order gradients.
However, the restrictions
(\ref{u-G.k})
imposed on the weak nonequilibrium ensure that
$|\Delta S^{(4)}| \ll |\Delta S^{(2)}|$.
For this reason, we conclude that the nonequilibrium entropy found
{\em is less\/}
than the entropy of the corresponding equilibrium state and as a
consequence it satisfies the second law of thermodynamics for
nonequilibrium processes
\cite{62GroMaz,70Gyarmati,83Bazarov.E}.

\section{Conclusions}
\label{Pidsumky}

We have considered the pressure, internal energy, and entropy of the
hard-sphere system in the weakly nonequilibrium heat-conduction steady
state.
They are calculated in the continuous media approach using integrations of
the proper local densities.

The results are obtained in the form of expansions in the temperature
gradients evaluated in the geometrical middle of the system up to the
fourth order.
They describe the effect of the particle size on thermodynamic
quantities at intermediate densities.
The coefficients of the expansions depend on the packing parameter
(referred to the uniform equilibrium),
revealing dependence of the nonequilibrium corrections on the volume of the
system.
The entropy calculated is shown to obey the second law of thermodynamics
for nonequilibrium processes.

The results are applicable for dimensions
$D = 1; 2; 3$.
The van der Waals approximation for hard spheres used restricts the
applicability to the domain of not high densities for
three- and two-dimensional systems where this approximation is valid for the
equilibrium.
In the one-dimensional case, the equilibrium van der Waals equation of state
is exact
\cite{36Tonks-pr}.
For this reason, we expect that our results can be used at high densities
while the probable inaccuracy may be caused only by the method used rather
than by the local equation of state.

Our calculations do not go beyond the scope of thermodynamic ideas,
since no external results coming from other nonequilibrium theories
(e.g., kinetic theory, informational theory, or the approach of fluctuation
theorems)
have been used.
The simplicity and explicit analytical description can be also regarded as
positive features.

\section*{Acknowledgement}

The author is grateful to Dr. T.M. Verkholyak for useful comments
and for reporting the work on the exact solution for the
equilibrium one-dimensional hard disks.

This research has made use of NASA's Astrophysics Data System.

\appendix

\section{Expressions for coefficients $\{ \nu \}$ and
$\{ \varkappa \}$.}
\label{nu.vk-g}

First of all, we expand quantities
$\{ \gamma \}$,
which define coefficients
$\{ \nu \}$
by the expressions given after
equation~(\ref{n.z-R}).
Note that
$\gamma_j$ is of the order
$\sim \delta^j$,
equation~(\ref{n.0-o}):
$$
\gamma_j =\frac 1 {j!} \,\frac {G_j} { T + bC^{(0)} }
\,\frac 1 { 1 + \delta^2 \zeta^{(2)} + \delta^4 \zeta^{(4)} + \ldots } \,,
$$
where
$\zeta^{(k)} \equiv b C^{(k)} / [T + bC^{(0)}]\big|_{k=2;\,4;\,\ldots}$.
For the fourth order it is sufficient to expand only
$\gamma_1$ and $\gamma_2$,
while
$\gamma_3$ and $\gamma_4$
are to be taken in the lowest approximation:
$$
\gamma_j \big|_{j=1;\,2} =
g_j \tilde\eta \big\{ 1 + \delta^2 [-\zeta^{(2)}] +\ldots \big\}
,\qquad
\gamma_j \big|_{j=3;\,4} =
g_j \tilde\eta \big( 1 + \ldots \big)
,$$
where
$g_j \equiv \tsf 1 {j!} G_j/T$.
A contribution from
$\zeta^{(2)}$
is also needed in
$\gamma_1^2
= g_1^2 \tilde\eta^2 \big\{ 1 + \delta^2 2[-\zeta^{(2)}] +\ldots \big\}$.

From the definitions given after
equation~(\ref{n.z-R}),
we find the lowest-order contributions to
$\{ \nu \}$ in the form:
$$
\begin{array}{ll}
\nu_1^{(0)} = - g_1 \tilde\eta
,&
\nu_3^{(0)} = - g_3 \tilde\eta + 2 g_2 g_1 \tilde\eta^2
- g_1^3 \tilde\eta^3
,\vspace{2mm}\\
\nu_2^{(0)} = - g_2 \tilde\eta + g_1^2 \tilde\eta^2
,\qquad\quad&
\nu_4^{(0)} =
- g_4 \tilde\eta + 2 g_3 g_1 \tilde\eta^2 + g_2^2 \tilde\eta^2
- 3 g_2 g_1^2 \tilde\eta^3 + g_1^4 \tilde\eta^4
.\end{array}
$$
The result
$ \zeta^{(2)} = - \tsf {L^2} {12} \tilde\eta^{-1} \eta \nu_2^{(0)}
$
gives the corrections as follows:
$$
\nu_1^{(2)} = \frac {L^2} {12} \eta
\left(g_2 g_1 \tilde\eta - g_1^3 \tilde\eta^2 \right)
,\qquad
\nu_2^{(2)} = \frac {L^2} {12} \eta
\left( g_2^2 \tilde\eta - 3 g_2 g_1^2 \tilde\eta^2 + 2 g_1^4 \tilde\eta^3 \right)
$$
which are of the third and fourth orders.
Due to the factor
$\eta \sim \overline n$,
$\nu_i^{(2)}$
are small in comparison with
$\nu_i^{(0)}$ at low densities.
Expressions for
$\varkappa_2^{(0)}$,
$\varkappa_2^{(2)}$, and
$\varkappa_4^{(0)}$
can be obtained using the formulae for
$\nu_i^{(k)}$ and
equations~(\ref{vkap-nu}).
Then, the rearranged coefficients read:
\bea
\nn
\varkappa_{(2)} &\eas=& \frac {L^2} {12}
\big( - g_2 \tilde\eta + g_1^2 \tilde\eta^2 \big)
,\\
\nn
\varkappa_{(4)} &\eas=& \frac {L^4} {80} \left[
- g_4 \tilde\eta + 2 g_3 g_1 \tilde\eta^2
+ g_2^2 \tilde\eta \left(1 - \frac 49 \eta\right)
- 3 g_2 g_1^2 \tilde\eta^2 \left(1 - \frac 49 \eta\right)
+ g_1^4 \tilde\eta^3 \left(1 + \frac 19 \eta\right)
\right]
.\eea

\newpage

\ukrainianpart

\title{Тиск і ентропія твердих кульок у слабонерівноважному
теплопровідному стаціонарному стані
}

  \author{Й.А. Гуменюк}
  \address{Інститут фізики конденсованих систем НАН України,
вул. Свєнціцького, 1, 79011 Львів, Україна
}

\makeukrtitle

\begin{abstract}
\tolerance=3000%
В рамках підходу суцільного середовища розраховано термодинамічні величини
системи твердих кульок у стаціонарному стані з малим тепловим потоком.
Аналітичні вирази для тиску, внутрішньої енергії та ентропії
знайдено в наближенні четвертого порядку за ґрадієнтами температури.
Показано, що ґрадієнтні внески до внутрішньої енергії залежать від об'єму, а
ентропія задовольняє
II-е начало термодинаміки для нерівноважних процесів.
Розрахунки проведено для вимірностей
3D, 2D та 1D.

\keywords тепловий потік, температурні ґрадієнти, рівняння стану, нерівноважна ентропія,
термодинаміка стаціонарного стану

\end{abstract}

\end{document}